# Formulating causal questions and principled statistical answers

Els Goetghebeur*[1] | Saskia le Cessie[2] | Bianca De Stavola[3] | Erica Moodie[4] | Ingeborg Waernbaum[5] | on behalf of the topic group Causal Inference (TG7) of the STRATOS initiative

[1]Department of Applied Mathematics, Computer Science and Statistics, Ghent University, Ghent, Belgium
[2]Department of Clinical Epidemiology/ Biomedical Data Sciences, Leiden Medical Center, Leiden, The Netherlands
[3]Great Ormond Street Institute of Child Health, University College London, London, U.K.
[4]Division of Biostatistics, McGill University, Montreal, Canada
[5]Department of Statistics, Uppsala University, Uppsala, Sweden

**Correspondence**
* Els Goetghebeur, Email: els.goetghebeur@ugent.be

**Summary**

Although review papers on causal inference methods are now available, there is a lack of introductory overviews on *what* they can render and on the guiding criteria for choosing one particular method. This tutorial gives an overview in situations where an exposure of interest is set at a chosen baseline ('point exposure') and the target outcome arises at a later time point. We first phrase relevant causal questions and make a case for being specific about the possible exposure levels involved and the populations for which the question is relevant. Using the potential outcomes framework, we describe principled definitions of causal effects and of estimation approaches classified according to whether they invoke the no unmeasured confounding assumption (including outcome regression and propensity score-based methods) or an instrumental variable with added assumptions. We discuss challenges and potential pitfalls and illustrate application using a 'simulation learner', that mimics the effect of various breastfeeding interventions on a child's later development. This involves a typical simulation component with generated exposure, covariate, and outcome data that mimic those from an observational or randomised intervention study. The simulation learner further generates various (linked) exposure types with a set of possible values per observation unit, from which observed as well as potential outcome data are generated. It thus provides true values of several causal effects. R code for data generation and analysis is available on www.ofcaus.org, where SAS and Stata code for analysis is also provided.

## 1 | INTRODUCTION

The literature on causal inference methods and their applications is expanding at an extraordinary rate. In the field of health research, this is fuelled by opportunities found in the rise of electronic health records and the revived aims of evidence-based precision medicine. One wishes to learn from rich data sources how different exposure levels *causally* affect expected outcomes in specific population strata so as to inform treatment decisions. While an abundance of machine learning techniques can handle electronic health records, they too need to integrate fundamental principles of causal inference to address causal questions.[1] Neither the mere abundance of data nor the use of a more flexible model pave the road from association to causation.

Experimental studies have the great advantage that treatment assignment is randomized. A simple comparison of outcomes on different randomized arms then yields an intention-to-treat effect as a robust causal effect measure. However, non-experimental or observational data remain necessary for several reasons. 1) Randomised controlled trials (RCTs) tend to be conducted in rather selected populations, to reduce costs and for ethical reasons. 2) We may seek to learn about the effect of treatments actually received in these trials, beyond the pragmatic effect of treatment assigned. This calls for an exploration of compliance with the assignment and hence for follow-up exposure data, i.e. non-randomized components of treatment received. 3) In many situations (treatment) decisions need to be taken also in the absence of RCT evidence 4) A wealth of patient data is being gathered in disease registries and other electronic patient records; these often contain more variables, larger sample sizes and greater population coverage than are typically available in an RCT setting. These needs and opportunities push scientists to seek causal answers in observational settings with larger and less selective populations, with longer follow-up, and with a broader range of exposures and outcome types (including adverse events).

Statistical causal inference has made great progress over the last quarter century, deriving new estimators for well-defined estimands using new tools such as directed acyclic graphs (DAGs) and structural models for potential outcomes.[2,3,4]

However, research papers – both theoretical and applied – tend to start from a question that is already formalised, and often describe published conclusions in vague causal terms missing a clear specification of the target of estimation. Typically, when this is specified, i.e. there is a well-defined estimand, a range of techniques can yield (asymptotically) unbiased answers under a specific set of assumptions. Several overview papers and tutorials have been published in this field. They are mostly focused, however, on the properties of one particular technique without addressing the topic in its generality. Yet in our experience, much confusion still exists about what exactly is being estimated, for what purpose, by which technique, and under what plausible assumptions. Here, we aim to start from the beginning, considering the most commonly defined causal estimands, the assumptions needed to interpret them meaningfully for various specifications of the exposure variable and the levels at which we might intervene to achieve different outcomes. In this way, we offer guidance on understanding what questions can be answered using various principled estimation approaches while invoking sensibly structured assumptions.

We illustrate concepts and techniques referring to a case study exemplified by simulated data, inspired by the Promotion of Breastfeeding Intervention Trial (PROBIT),[5] a large RCT in which mother-infants pairs across 31 Belarusian maternity hospitals were randomised to receive either standard care or an offer to follow a breastfeeding encouragement programme. Aims of the study were to investigate the effect of the programme and breastfeeding on a child's later development. We use simulated data to examine weight achieved at age 3 months as the outcome of interest in relation to a set of exposures defined starting from the intervention and several of its downstream factors. Our simulation goes beyond mimicking the 'observed world' by also simulating for every study participant how different exposures strategies would lead to different potential responses. We call this the *simulation learner* PROBITsim and refer to the setting as the Breastfeeding Encouragement Programme (BEP) example. Source code for implementation is available on www.ofcaus.org.

Our aim here is to give practical statisticians a compact but principled and rigorous operational basis for applied causal inference for the effect of point (i.e. baseline) exposures in a prospective study. We build up concepts, terminology, and notation to express the question of interest and define the targeted causal parameter. In section 2, we lay out the steps to take when conducting this inference, referring to key elements of the data structure and various levels of possible exposure to treatment. Sections 3 presents the potential outcomes framework with underlying assumptions, and formalises causal effects of interest. In section 4, we describe PROBITsim, our simulation learner. We then derive various estimation approaches under the no unmeasured confounding assumption and under the instrumental variable assumption in section 5. We explain how the approaches can be implemented for different types of exposures, and apply the methods on the simulation learner in section 6. We end with an overview that highlights overlap and specificity of the methods as well as their performance in the context of PROBITsim, and more generally. R code for data generation, R, SAS and STATA code for analysis and reporting, and slides that accompany this material and apply the methods to a second case study can be found on www.ofcaus.org.

## 2 | FROM SCIENTIFIC QUESTIONS TO CAUSAL PARAMETERS

Causal questions ask what would happen to outcome $Y$, had the exposure $A$ been different from what is observed. To formalise this, we will use the concept of potential outcomes[6,7] that captures the thought process of *setting* the treatment to values $a \in \mathcal{A}$, a set of possible treatment values, without changing any pre-existing covariates or characteristics of the individual. Let $Y_{\mathfrak{a}(a)}$ be the potential outcome that would occur if the exposure were set to take the value $a$, with notation $\mathfrak{a}(a)$ indicating the action of *setting A* to *a*. In what follows we will refer to *A* as either an 'exposure' or a 'treatment' interchangeably. Since individual level causal effects can never be observed, we focus on expected causal contrasts in certain populations. In the BEP example there are several linked definitions of treatment; these include 'offering a BEP', 'following a BEP', 'starting breastfeeding' or 'following breastfeeding for 3 full months'. Each of them may require a decision of switching the treatment on or off. Ideally this decision is informed by what outcome to expect following either choice.

It is important that causal contrasts should reflect the research context. Hence in this example one could be interested in evaluating the effectiveness of the programme for the total population or in certain sub-populations. However, for some sub-populations the intervention may not be suitable and thus assessing causal effects in such sub-populations would not be useful.

Consider the following question: *"Does a breastfeeding intervention, such as the one implemented in the PROBIT trial, increase babies weight at three months?"* Despite its simplicity, empirical evaluation of this question involves its translation into meaningful quantities to be estimated. This requires several intermediate steps:

1. Define the treatment and its relevant levels/values corresponding to the scientific question of the study.

2. Define the outcome that corresponds to the scientific questions under study.

3. Define the population(s) of interest.

4. Formalise the potential outcomes, one for each level of the treatment that the study population could have possibly experienced.

5. Specify the target causal effect in terms of a parameter, i.e. the *estimand*, as a (summary) contrast between the potential outcome distributions.

6. State the assumptions validating the causal effect estimation from the available data.

7. Estimate the target causal effect.

8. Evaluate the validity of the assumptions and perform sensitivity analyses as needed.

Explicitly formulating the decision problem one aims to solve or the hypothetical target trial one would ideally like to conduct[8] may guide the steps outlined above. In the following we expand on steps 1-5 before introducing the simulation learner in Section 3 and discussing steps 6-8 in Section 4.

## 2.1 | Treatments

Opinions in the causal inference literature differ on how broad the definition of treatment may be. Some say that the treatment should be manipulable, like giving a drug or providing a breastfeeding encouragement programme.[9] Here, we take a more liberal position which would also include for example genetic factors or even (biological) sex as treatments. Whichever the philosophy, considered levels of the treatments to be compared need a clear definition, as discussed below.[10]

Treatment definitions are by necessity driven by the context in which the study is conducted *and* the available data. The causal target may thus differ for a policy implementation or a new drug registration, for instance, or whether the data are from an RCT or administrative data. In the BEP example we may wish to define the causal effect of a breastfeeding intervention on the babies' weight at three months. There are several alternative specifications of a 'breastfeeding treatment' possible. Below we list a few which are interconnected and represent different types of treatment decisions:

- $A_1$ : (randomised) treatment prescription: e.g. an encouragement programme was offered to pregnant women.

- $A_2$ : uptake of the intervention: e.g. the woman participated in the programme (when offered).

- $A_3$ : uptake of the target of intervention: e.g. the mother started breastfeeding.

- $A_4$ : completion of the target of intervention: e.g. the mother started breastfeeding and continued for three months.

Each of these treatment definitions $A_k$, $k = 1, ..., 4$, refers to a particular breastfeeding event taking place (or not). A public health authority will be more interested in $A_1$ because it can only decide to offer the BEP or not; an individual mothers interste will be in the effect of $A_2$, $A_3$ and $A_4$ because she decides to participate in the programme and to start and maintain breastfeeding. For any one, several possible causal contrasts may be of interest and are estimable. See Section 2.6.

It is worth noting that these various definitions are not all clear-cut. For example, while $A_4 = 1$ may be most specific in what it indicates, $A_4 = 0$ represents a whole range of durations of breastfeeding: from none to 'almost 3 months'. In the same vein, $A_3 = 1$ represents a range of breastfeeding durations that follow initiation, against $A_3 = 0$ which implies no breastfeeding at all. The variation in underlying levels of treatment could be seen as multiple versions of the treatment; we consider this topic further in Section 3.2.

Intervening at a certain stage in the 'exposure chain' likely affects downstream exposure levels, as reflected in Figure 1. This is the set-up we have used to generate the simulation learner dataset (see Section 3), with the BEP being only available to those randomized, and where uptake of the programme increases the probability of $A_3 = 1$ and, importantly, also increases breastfeeding duration among women who initiate breastfeeding. There are of course many further aspects of the breastfeeding process that could be considered when defining exposures that are downstream from an initial randomized intervention, e.g. maternal diet, the timing and frequency of breastfeeding, exclusive versus predominant breastfeeding, and so on, however for didactic purposes, we shall omit such considerations.

## 2.2 | Outcomes

Similar to the definition of the treatment, it is important to carefully define the outcome $Y$. In the BEP example, the outcome of interest could be the infant's weight at three months, or the increment between birth weight and weight at three months or whether the infant is above a certain weight at 3 months. Typically, the distribution of both the absolute and relative weight are of interest: a BEP may well increase mean weight at 3 months by 200 grams but also increase the number of overweight infants. Clarity of which outcome definition corresponds to the question of interest is therefore crucial.

## 2.3 | Populations

A causal effect will in most cases vary between populations due to effect modification. It is then important to identify and describe the population to whom a stated effect pertains. Often one is interested in the causal effects in several sub-populations. Researchers and policy-makers might want to study if the breastfeeding intervention is substantially more effective for infants of less educated women who may be at highest risk of being born low weight. They could also wonder about the effect of treatment in the subpopulation of those who are actually exposed (the 'treated', as discussed above).

In the next section we will develop causal effects for the different sub-populations. In most settings we want to consider populations of individuals who have the possibility of receiving all treatment levels of interest. This condition is referred to as *the positivity assumption*.[11] It could be violated, for example, if the target population included women for whom breastfeeding is precluded (because of pre-existing or pregnancy-related conditions). Studying the effect of breastfeeding in the subpopulation of infants whose mothers cannot breastfeed (or indeed a larger population that includes this subgroup) may be impossible due to missing information – and indeed irrelevant.

## 2.4 | Potential outcomes

As stated above, a potential outcome $Y_{\mathfrak{a}(a)}$ is the outcome we would observe if an exposure were *set* at a certain level $a$, where $\mathfrak{a}(a)$ indicates the action of *setting* $A$ to $a$. This notion needs some additional considerations linking it to the treatments and outcomes definitions given above. Specifically there are two commonly invoked assumptions that help achieve this: *no interference* and *causal consistency*.

### *No interference*

No interference means that the impact of treatment on the outcome of individual $i$ is not altered by other individuals being exposed or not. At first sight this is likely justified in our setting: one baby's weight typically does not change because another baby is being breastfed. In resource poor or closely confined settings this could, however, be challenged. For instance, interference would happen when a child is affected by the consequences of a reduced immune system of other children who were not breastfed and hence became more susceptible to infectious diseases which may impact their weight at three months.

When the assumption of no interference is not met, the potential outcome definition becomes much more complex and involves the treatment assigned to other individuals.[12] For example, if there were interference among infants living in the same household, the potential outcome of infant $i$ would be defined not as $Y_{\mathfrak{a}(a)}$ but as $Y_{\mathfrak{a}_i(a),\mathfrak{a}_{i_1}(a^*),...,\mathfrak{a}_{i_{K_i}}(a^\dagger)}$, where infants $i_1$ to $i_{K_i}$ belong to the same household as infant $i$ and their breastfeeding status is set to take values $(a^*, ..., a^\dagger)$.

*Causal consistency*

The assumption of causal consistency relates the observed outcome to the potential outcomes. Consistency (at an individual level) means that $Y_{\mathfrak{a}(a)} = Y$ when $A = a$, hence assuming consistency implies that the observed outcome in our data is the same as the potential outcome that would be realised in response to setting the treatment to the level of the exposure that was observed. This directly affect our interpretation of the estimated causal effect for the study population. It will also affect transportability to new settings in ways that may be hard to predict.

In practice this implies that the mode of receiving as opposed to choosing treatment level $A = a$ per se has no impact on outcome. This may not be the case for many real life settings. For example 'starting breastfeeding' ($A_3 = 1$) potentially has multiple versions as some mothers who initiate breastfeeding may continue to do so for at least 3 months, while others may discontinue sooner. Also, breastfeeding may be exclusive or supplemented, breast milk may be fed at the breast or with a bottle, and so on. Hence it is to be expected that setting $A_3$ to be 1 may translate into different durations and types of breastfeeding, and thus may not lead to the same infant weight at 3 months as when starting breastfeeding is a choice. More generally, it is typically the case that a treatment can come in many variations at some level of resolution. To achieve consistency then a more precise definition of treatment is required, so that observing or setting it is more likely to generate comparable effects.

These observations relate to the importance of a well-defined exposure[13] and the need to be as precise as the data allow us in our definition of treatment.[14] Some authors have criticised the restriction imposed by this assumption (and hence by the potential outcomes approach to causal inference[10]). Being aware of the possibility of multiple versions of treatment should not deter us from pursuing the most relevant definition of treatment: instead it should lead us to greater precision and transparency in formulating the causal question and its transportability.

Note also that the assumption of consistency may be relaxed by rephrasing it at the distributional level (possibly conditional on baseline covariates), in the sense that consistency would concern for example the equality of the mean observed outcome of those with observed values $A = a$ and the mean potential outcome had their treatment been set to $a$. Following this broader definition, any causal interpretation would be applicable only to settings where the *distribution* of the different versions of treatment equaled that in the analyzed sample.

## 2.5 | Nested potential outcomes

The treatments considered here belong to a chain: when $A_1$ is set, it has consequences for the "worlds" where $A_2$, $A_3$ and $A_4$ act. Equivalently, when $A_3$ is set at a subsequent baseline time, where breastfeeding starts, $A_1, A_2$ become baseline covariates with consequences for the worlds that follow. Although this paper does not enter into the full framework of estimation for dynamic treatment strategies, we can benefit from additional definitions of potential outcomes that recognise the nested nature of the interventions.

Below we define worlds where setting $A_2$ and $A_3$ occurs under alternative scenarios that depend on how $A_1$ was set (and, for $A_3$, how $A_1$ and or $A_2$ was set). These will be useful for the discussion in 2.6.

1. In the world where BEP is on offer to all (i.e. when $\mathfrak{a}_1(1)$ is set for everyone in the population), the potential outcomes of participating or not participating in the BEP are defined as $Y_{\mathfrak{a}_1(1),\mathfrak{a}_2(1)}$ and $Y_{\mathfrak{a}_1(1),\mathfrak{a}_2(0)}$. Similarly in the world where BEP is not offered, we may consider the potential outcome of not participating in the BEP defined as $Y_{\mathfrak{a}_1(0),\mathfrak{a}_2(0)}$. In our example we assumed that the programme was only available to the intervention group (i.e. $Y_{\mathfrak{a}_1(0),\mathfrak{a}_2(1)}$ is not defined), and that the intervention would only affect outcome if the programme was actually followed (i.e. $Y_{\mathfrak{a}_1(1),\mathfrak{a}_2(0)} = Y_{\mathfrak{a}_1(0),\mathfrak{a}_2(0)}$). (In other settings it is conceivable that the mere invitation to BEP, comes with advice that may have a direct impact on outcome under $\mathfrak{a}_2(0)$).

2. Setting $\mathfrak{a}_2(1)$, here implies that $A_1$ is set to 1; setting $\mathfrak{a}_2(0)$ can happen independently of how $A_1$ is set. The corresponding potential outcomes are denoted by $Y_{\mathfrak{a}_2(1)}(= Y_{\mathfrak{a}_1(1),\mathfrak{a}_2(1)})$ and $Y_{\mathfrak{a}_2(0)}$.

3. Similarly, when interest is in the causal effect of $A_3$, the potential outcomes of starting or not starting breastfeeding in the world with BEP on offer are $Y_{\mathfrak{a}_1(1),\mathfrak{a}_3(1)}$ and $Y_{\mathfrak{a}_1(1),\mathfrak{a}_3(0)}$, and in the world without BEP, they are $Y_{\mathfrak{a}_1(0),\mathfrak{a}_3(1)}$ and $Y_{\mathfrak{a}_1(0),\mathfrak{a}_3(0)}$. We deliberately omitted setting/fixing the possible $\mathfrak{a}_2$ level here, because we let it follow the natural course after setting $\mathfrak{a}_1(1)$, meaning that women may or may not choose to follow the BEP, after receiving the offer. The effect of breastfeeding in the world where the BEP is offered, may differ from the effect when the BEP is not available, as the BEP may not only affect the probability to start breastfeeding, but also the duration of breastfeeding for those who start.

4. One could be tempted to evaluate $Y_{\mathfrak{a}_3(1)}$ in the study context, ignoring $A_1$ and effectively averaging over the observed $A_1$, distribution which follows from equal randomization. Such random BEP offer is however no intended real world future scenario and hence that particular average effect measure is of no direct relevance to any realistic scenario.

5. In the world where all women follow the programme (i.e. $\mathfrak{a}_2(1)$ is set, implying also $\mathfrak{a}_1(1)$ as we assume BEP cannot be followed unless it is offered), the potential outcomes of starting breastfeeding or not are $Y_{\mathfrak{a}_2(1),\mathfrak{a}_3(1)}$ and $Y_{\mathfrak{a}_2(1),\mathfrak{a}_3(0)}$.

## 2.6 | Causal parameters

The next step is to contrast potential outcomes under different settings of exposure variables. We do so by defining an estimand in a well-defined (sub)population. Individual causal effects cannot be computed since each individual can only be assigned to one treatment at a time since via consistency one and only one potential outcome can be observed. However, population summary measures can be estimated (under additional assumptions to be discussed below) for different groups, such as the total population or the sub-population of treated (or untreated) individuals. Also, causal effects can be defined on different scales. In this paper we focus on the mean difference as the contrast of interest.

Table 1 describes a selection of causal parameters for exposures $A_1$ and $A_2$. The first estimand for $A_1$ listed in the table is the average treatment effect in the population ($ATE_1$) and corresponds to the question *"What would the average infant weight at 3 months be had all mothers been offered the BEP, versus the average infant weight had the mothers not been offered the programme?"*. It is defined as $ATE_1 = E[Y_{\mathfrak{a}_1(1)}] - E[Y_{\mathfrak{a}_1(0)}]$, which is equal to the intention to treat effect(ITT) of the randomised trial.

There are several possible contrasts involving uptake of the intervention $A_2$. We could target the causal question *"What would the average infant weight at 3 months be had all mothers attended the BEP, versus the average infant weight had none of the mothers attended the programme?"* over the whole infant population, leading to $ATE_2 = E[Y_{\mathfrak{a}_2(1)}] - E[Y_{\mathfrak{a}_2(0)}]$. We might also consider this effect only within the population of women who chose to accept the offer and did attend the BEP. The latter would be the ATT. Because in our example the BEP is only available to those who are offered it, the treated population are those with $A_2 = 1$ and $A_1 = 1$; see Table 1 . The effect in the population, $ATE_2$ would be of overall interest to the developers of the BEP, as would the tratment effect in the non treated $ATNT_2$, because the latter would quantify the gain to be expected from a more convincing promotion campaign for the current programme with larger attendance, i.e. a greater $P(A_2 = 1 | A_1 = 1)$. In contrast $ATT_2$ might be of greater interests to mothers following BEP, as this would provide a measure of the expected benefit from their own uptake of the BEP offer.

Furthermore, causal effects may be heterogeneous across observable strata, for instance if the breastfeeding treatment has different causal effects depending on the education level of the mother. Thus causal effects specific to baseline subgroups would be of interest, e.g. the average causal effect among those with low education and compare it to the average causal effect among those with high education. We can also define a causal effect conditional on multiple characteristics such as the expected causal effect of the programme in the group of 30 year old smoking mothers with a child born by caesarian section.

## 3 | THE SIMULATION LEARNER

To illustrate concepts and support our learning, we generated data informed by a real investigation but enriched by the generation of potential outcome data in addition to 'observed' data. We follow Wallace et al.[15] in simulating data to mimic the results of the Promotion of Breastfeeding Intervention Trial[5] (PROBIT). Mother-infant pairs were randomised to receive either standard care or a breastfeeding encouragement intervention. In these simulations we are focussing on weight achieved at age 3 months, thus the study population consists of babies who survive the first three months.

## 3.1 | Generating the variables

Figure 1 outlines the main relationships among the simulated variables. The baseline variables $L_1$ were: mother's age, location of living (urban versus rural and western versus eastern region), level of education (low, intermediate, high), maternal history of allergy, and smoking during pregnancy. The variables related to the infant's birth $L_2$ were: sex of child, birth weight, and birth by caesarian section. Thus, $L_1$ are confounders of the relationship between $A_2$ and $Y$ , and $(L_1, L_2)$ are confounders of the relationship between $A_3$ and $Y$. The distribution of these variables was made to resemble that of the PROBIT study and the sample size $n$ was set to 17,044, as in that study. Details of the data generation process can be found in Appendix 1 and in the material available at www.ofcaus.org; an overview is given below.

The offer of the programme ($A_1$) was assigned randomly, but the uptake of programme ($A_2$), starting breastfeeding ($A_3$), and the duration of breastfeeding ($A_4$) were all affected by variables at baseline ($L_1$) or at birth ($L_2$), with their union denoted by the vector $L$. We made the simplifying assumptions that $L_2$ were unaffected by the programme offer, that the programme was only available to women in the intervention group, and that the intervention would only affect outcome if the programme was actually followed. The odds of following the programme after receiving an offer was assumed to depend on maternal age, education and smoking during pregnancy, such that older and more highly educated women had a higher probability of following the programme, while smokers were less likely to do so.

Following the programme, i.e. $A_2 = 1$ was set to influence weight at 3 months in two ways: it increased the probability of starting breastfeeding, and increased the duration of breastfeeding. Older and more highly educated women and women who did not smoke during pregnancy were more likely to start breastfeeding, while having a child with lower birth weight or a baby girl decreased the probability of starting breastfeeding. The uptake of the program, higher age, higher education, not smoking, a higher birth weight, and maternal allergies were set to increase the total duration of breastfeeding, while delivery by caesarian or a having baby boy to lower it.

The outcome (weight at 3 months) was set to be affected by the duration of breastfeeding and by the baseline and birth variables, some of which (smoking, education and birth weight) also modified the effect of breastfeeding.

For each woman in the simulated dataset, we observed realised values of $A_1$, $A_2$, $A_3$ and $A_4$ and of the weight of the child after 3 months. In addition several potential outcomes were generated representing the potential weight at 3 months of the child under different interventions on $A_1$, $A_2$, $A_3$ and $A_4$. This means that in our dataset, for each woman the potential weight of her child at 3 months is known under different scenarios: if she had received the offer for the BEP, if she had not received the offer, if she had followed the programme, if she had or had not started breastfeeding and if she had continued breastfeeding for 3 months. Our simulations generated correlated potential outcomes, but the causal parameters introduced so far are not affected by this. We see this as an advantage since there is an intrinsic lack of information on the joint distribution of the potential outcomes in observed data. Table 2 gives the expected value of the different potential outcomes overall and in specific strata (subpopulations). These values were obtained from a very large simulated dataset of 5 million observations and are here considered to represent the truth.

## 3.2 | Different causal contrasts

From Table 2 we can derive several true causal contrasts. For example the average treatment effect (ATE) of the BEP offer is $ATE_1 = E[Y_{\mathbf{a}_1(1)}] - E[Y_{\mathbf{a}_1(0)}] = 6115 - 6017 = 98$ grams. This effect may be of interest to policy makers as it is the overall mean change in infant weight at 3 months due to inviting expectant women to attend the BEP. Comparing the scenario where everyone actually receives the offer and follows the BEP with no programme, the expected weight gain is $ATE_2 = E[Y_{\mathbf{a}_2(1)}] - E[Y_{\mathbf{a}_2(0)}] = 165$ grams. Among women who actually follow the programme (the treated), the effect of BEP uptake is $ATT_2 = E[Y_{\mathbf{a}_2(1)}|A_2 = 1] - E[Y_{\mathbf{a}_2(0)}|A_2 = 1] = 153$ grams. The effect of participating in the BEP among women who have the opportunity to follow it but opt not to, is $ATNT_2 = E[Y_{\mathbf{a}_2(1)}|A_2 = 0, A_1 = 1] - E[Y_{\mathbf{a}_2(0)}|A_2 = 0, A_1 = 1] = 185$ grams. $ATNT_2$ is larger than $ATT_2$ because women who would benefit most from the BEP were, in our simulated dataset, less inclined to follow it.

In this tutorial, we are treating $A_1$, $A_2$, $A_3$ and $A_4$ as point exposures, i.e. as exposures to be examined separately, with any previous exposures in the chain treated as background variables. In other words, for each targeted treatment, we consider the time point at which it is implemented. We then ask about the impact of setting this treatment to a given value, conditional on background information. In the setting of our study: when $A_3$, the decision to start breastfeeding is implemented, the values of $A_1$ and $A_2$ are already known and the baby has been born. The set of information carried by $A_1$ and $A_2$ could be treated as baseline information, like $L$, conditional on which the effect of starting breastfeeding is measured.

Alternatively, we could consider the joint impact of multiple interventions. Using the nested potential outcomes notation introduced in section 2.5, we could address the question *"What would the average infant weight at 3 months be had all mothers started breastfeeding versus the average infant weight had they not started at all?"* under different worlds where $A_1$ and $A_2$ are set to take different values. In the world without BEP, the answer would be $ATE_{3,\mathbf{a}_1(0)} = E[Y_{\mathbf{a}_1(0),\mathbf{a}_3(1)}] - E[Y_{\mathbf{a}_1(0),\mathbf{a}_3(0)}] = 387$ grams. In the world where the BEP is offered, the gain in weight at 3 months would be substantially higher: $ATE_{3,\mathbf{a}_1(1)} = E[Y_{\mathbf{a}_1(1),\mathbf{a}_3(1)}] - E[Y_{\mathbf{a}_1(1),\mathbf{a}_3(0)}] = 422$ grams. The weight gain in the world where everyone followed the programme would be $ATE_{3,\mathbf{a}_2(1)} = Y_{\mathbf{a}_2(1),\mathbf{a}_3(1)} - Y_{\mathbf{a}_2(1),\mathbf{a}_3(0)} = 450$ grams. This is the largest effect because, in the simulation, BEP increases the mean duration of breastfeeding. In general there are greater average potential outcomes with increased intensity of the joint interventions.

The average treatment effect in the treated (with respect to $A_3$) also differs between randomisation worlds because more women among those randomised to receive the BEP will start breastfeeding than in the control group. The effect of breastfeeding in those who started breastfeeding and are in the intervention arm (i.e. $A_1 = 1$) is equal to $\text{ATT}_{3, \mathbf{a}_1(1)} = E[Y_{\mathbf{a}_1(1), \mathbf{a}_3(1)} | A_3 = 1, A_1 = 1] - E[Y_{\mathbf{a}_1(1), \mathbf{a}_3(0)} | A_3 = 1, A_1 = 1] = 421$ grams, and the effect of breastfeeding in those who started breastfeeding but are in the control arm (i.e. $A_1 = 0$) is $\text{ATT}_{3, \mathbf{a}_1(0)} = 381$ grams. The average effect of breastfeeding in those who did not start breastfeeding is $\text{ATNT}_{3, \mathbf{a}_1(1)} = 424$ gram when the programme is available and $\text{ATNT}_{3, \mathbf{a}_1(0)} = 393$ grams when not.

We could also ask the question *"What would the average infant weight at 3 months be, had all mothers breastfeed for 3 months versus the average infant weight had they not started at all?"* As noted before, setting $A_4 = 0$ will include a very heterogeneous set of breastfeeding behaviours, as well as not breastfeeding at all. A more refined question would restrict the comparison to a setting where there is no breastfeeding at all, i.e. $E[Y_{\mathbf{a}_4(1)}] - E[Y_{\mathbf{a}_3(0)}] = 6351 - 5827 = 524$ grams.

When implementing an intervention, it is of interest to identify those subgroups for which the intervention is most beneficial. Table 2 for example, shows that the infants of mothers in the lowest stratum of education would gain more than those of mothers in the highest, both when the intervention is offering the programme $E[Y_{\mathbf{a}_1(1)} | L = \text{low}] - E[Y_{\mathbf{a}_1(0)} | L = \text{low}] = 110$ grams and when the intervention is following the programme $E[Y_{\mathbf{a}_2(1)} | L = \text{low}] - E[Y_{\mathbf{a}_2(0)} | L = \text{low}] = 214$ grams, as opposed to 66 grams and 85 grams for women in the highest stratum of education.

Some of the causal effects described above are not realistic. For example, the largest causal contrast is the expected weight gain when every infant is breastfed for the full 3 months versus the expected weight gain when no one is breastfed (524 grams above). However not all women can or wish to start breastfeeding (nor would all women willingly refrain from it). As alluded to in the discussion of positivity in Section 2.3, a woman who is very ill at the end of pregnancy may not have the option of breastfeeding her baby because of toxicity of prescribed medication or ill-health. It follows that considering the intervention where every woman continues breastfeeding for the full 3 months is even less realistic. It is important to define the causal question precisely in a pertinent population before turning to estimation.

## 4 | PRINCIPLED ESTIMATION APPROACHES

The estimation approaches discussed here rely on further assumptions in addition to those outlined in section 2.4. These can be classified according to whether or not they invoke the *no unmeasured confounding* (NUC) assumption which states that the received treatment is independent of the potential outcomes, given covariates $\mathbf{L}$. Formally, the NUC assumption states: $(Y_{\mathbf{a}(0)}) \perp A | \mathbf{L}$ and $(Y_{\mathbf{a}(1)}) \perp A | \mathbf{L}$, where, hereafter $A$ denotes a binary exposure. In other words, the assumptions states that a sufficient set of variables $\mathbf{L}$ that confound the exposure/outcome relationship have been measured and are available to the analyst.

The estimation approaches that rely on the NUC assumption include standard outcome regression and propensity score (PS) based methods such as PS stratification, regression, matching, and inverse probability weighting. These are reviewed below. Alternatively, if an instrumental variable (IV) is available, IV methods can be used by also invoking additional assumptions in place of NUC. IV definitions an assumptions are described in Section 4.2.

### 4.1 | Methods based on the No Unmeasured Confounders assumption

When a sufficient set of confounders $\mathbf{L}$ is measured, the causal effect of treatment can be estimated by comparing observed outcomes between the treated and untreated people with identical values for $\mathbf{L}$. Such direct control for $\mathbf{L}$ may be done in different ways: by regression or stratification or matching. We discuss these approaches in the next subsections.

Before proceeding with the analysis one should examine how treatment groups differ in their population mix – that is, examine the imbalance in covariates between treatment groups. This information can then be summarised in the *propensity score*.

The propensity score (PS) is the probability of being treated conditional on the covariates, $e(\mathbf{L}) = P(A = 1 | \mathbf{L})$.[16] The PS is an important function of the covariates that reduces the (possibly high dimensional) vector $\mathbf{L}$ into a scalar containing all information that is relevant for the treatment assignment in relation to the outcome. This property is called the balancing property, meaning that the covariate distributions of the treated and non-treated are the same when conditioning on the PS. Intuitively, the role of the PS can be thought of as one of restoring balance between treated and untreated groups. For example, if we were to compare all treated subjects with untreated subjects who all had the same value of the PS, the distribution of the

covariates $L$ would be the same, much like in a randomized trial. However unlike in a randomized trial, balance is not achieved between the treated and untreated groups for any covariates that were not included in the PS. The balancing property implies that all relevant confounding information in $L$ is contained in $e(L)$, so that if $(Y_{\mathfrak{a}(0)}, Y_{\mathfrak{a}(1)}) \perp A|L$, then also $(Y_{\mathfrak{a}(0)}, Y_{\mathfrak{a}(1)}) \perp A|e(L)$. This implies that $e(L)$ can be used instead of the full vector $L$.

The PS is estimated from the data, usually by fitting a parametric (e.g. logistic regression) model for the probability of being treated given the confounding variables, although a variety of other approaches can be employed including tree-based classification.[17] However derived, the adequacy of the estimated PS, $\hat{e}(L)$, as a balancing summary of the confounder distributions across treatment groups must be evaluated[18] by checking whether $L \perp A \mid \hat{e}(L))$. While balance of the joint distribution of the confounders $L$ is required, in practice balance is often assessed for each confounder $L \in L$ separately by comparing standardized mean differences, variance ratios, and other distributional statistics and plots such as empirical cumulative distribution plots, between the treated and untreated groups after weighting, stratification, or matching by the estimated propensity score.[19] We illustrate some of these checks in Appendix 2. To date, variable selection for propensity score modelling is done largely on a trial and error basis, beginning with a model thought to contain all relevant confounders and adding higher order terms (polynomials, interactions) if balance appears not to have been achieved.

The propensity score can also be used to examine the positivity assumption by checking for overlap of the propensity score distribution of those who are treated and those who are not. For this reason, automatic variable selection approaches (e.g. stepwise) or prediction-based measures of fit (e.g. C-statistic), which seek best prediction of treatment allocation when specifying the PS model, may not provide the best balance for the confounders and favour variables that are strongly predictive of the treatment, even if they are only weakly or not at all predictive of the outcome.

### 4.1.1 | Outcome regression

Perhaps the simplest and most familiar form of causal estimation is outcome regression. In this approach, a model is posited for the outcome as a function of the exposure and the covariates. For example, for a continuous outcome the linear regression model of the form

$$E[Y|A, L] = \beta_0 + \beta_1 A + \beta_2' f(L, A)$$

where $\beta_2$ is a vector of parameters and $f(L, A)$ is a (vector) function of $L$ and $A$ representing, for example, the main effect of the covariates $L$ and interactions between covariates and $A$. Ordinary least squares can be used to estimate the parameters of the outcome linear regression model.

Assuming no interference, consistency and NUC, $\beta_1$ is interpreted as the causal effect of $A$ in the reference categories of $L$ (i.e. where $L = 0$ if $f(L, A)=0$ when $L = 0$). This is a conditional causal effect.

To estimate causal parameters such as those shown in Table 1 , the additional step of marginalizing over the distribution of $L$ is needed. For example, assuming that the model below is correctly specified

$$E[Y|A, L] = \beta_0 + \beta_1 A + \beta_2' L + \beta_3' LA$$

we identify the ATE for $A$ as follows:

$$
\begin{aligned}
\text{ATE} &= E\{E[Y_{\mathfrak{a}(1)}|L]\} - E\{E[Y_{\mathfrak{a}(0)}|L]\} \\
&= E\{E[Y_{\mathfrak{a}(1)}|A = 1, L]\} - E\{E[Y_{\mathfrak{a}(0)}|A = 0, L]\} \\
&= E\{E[Y|A = 1, L]\} - E\{E[Y|A = 0, L]\} \\
&= (\beta_0 + \beta_1 + \beta_2' E[L] + \beta_3' E[L]) - (\beta_0 + \beta_2' E[L]) \\
&= \beta_1 + \beta_3' E[L],
\end{aligned}
$$

where the second equality follows from the NUC assumption, the third from the consistency assumption and the fourth from the assumption of correct specification of the outcome model. This estimand can be estimated by $\hat{\beta}_1 + \hat{\beta}_3' n^{-1} \sum_{i=1}^{n}(l_i)$, where $n$ is the sample size. When there are no treatment-covariate interactions (i.e. $\beta_3$ is a vector of zeroes), then the ATE equals $\beta_1$ and its standard error can be taken directly from the fitted model that does not include any interactions. Otherwise, a standard error accounting for the correlation between $\beta_1$ and $\beta_3$ as well as estimation of $E[L]$ must be computed either analytically or via a bootstrap procedure.

A similar approach can be taken to estimate the ATT (or the ATNT). The ATT, for instance, can be computed noting that ATT $= E\{E[Y_{\mathfrak{a}(1)}|A = 1, L]\} - E\{E[Y_{\mathfrak{a}(0)}|A = 1, L]$. Letting $\mathcal{I}_{A=1}$ denote the indices $i$ of those exposed subjects and

$\#\mathcal{I}_{A=1} = \sum_{i=1}^{n} a_i$ denote the number of exposed individuals (the cardinality of $\mathcal{I}_{A=1}$), the ATT can be estimated using the outcome regression coefficient estimates by

$$\widehat{\text{ATT}} = (\#\mathcal{I}_{A=1})^{-1} \sum_{i \in \mathcal{I}_{A=1}} (\hat{\beta}_1 + \hat{\beta}_3' l_i).$$

For binary and other categorical outcomes other appropriate outcome models can be used such as the logistic regression model. This model will yield fitted values of $E[Y|A = 1, L]$ and $E[Y|A = 1, L]$ for all individuals which can then be averaged over the appropriate population.

Concerns about model mis-specification may be reduced by using a more flexible model for the outcome. For example, we may consider transformations of $L$ such as splines to specify $f(L, A)$, leading to a less parametric model which, however, requires estimation of a greater number of parameters. An additional concern, is the possibility that a chosen outcome model leads to extrapolations outside of the data cloud (in other words, to lack of positivity). Users should therefore be aware of this and adopt methods discussed above to assess whether lack of positivity is an issue.

When an appropriate propensity score has been estimated such that it provides the desired balance, outcome regression can also be performed with the generic function $f(L, A)$ being replaced by $\hat{e}(L)$, assuming no interactions between $L$ and $A$:

$$E[Y|A, L] = \beta_0 + \beta_1 A + \beta_2 \hat{e}(L).$$

This approach is known simply as propensity score regression with the ATE and ATT then estimated via standard regression followed by averaging over the PS as opposed to $L$, much as in Section 4.1.1. It can be shown that for the linear outcome model the propensity score regression estimators for the ATE and ATT are consistent under correct specification of the propensity score, even if the outcome model is mis-specified, provided the treatment effect is constant across $e(L)$.[20]

### 4.1.2 | Stratification and matching

Stratification can be used to estimate the ATE by taking the weighted sum of the differences of sample means across strata, where strata may be defined by ranges of the propensity score (fifths – i.e. using quintiles – is a common choice,[21] for large sample sizes increasing the number of strata will reduce the residual bias within strata)). The same approach could be used by stratification by the covariates $L$, although this may be more challenging than stratifying or matching on the PS if $L$ is not low-dimensional.

Let $n_j$ denote the number of individuals in stratum $j = 1, ..., J$. Finally, let $\hat{\mu}_{aj}$ denote the sample average of $Y$ for those with treatment level $a$ in the $j$-th stratum. Then the stratification-based estimator of the ATE is given by

$$\sum_{j=1}^{J} \left( \frac{n_j}{n} \right) [\hat{\mu}_{1j} - \hat{\mu}_{0j}].$$

This approach will work if there is reasonable balance of values of confounders in each of the defined strata. If not, one can regress the outcome on confounders within strata and use the stratum-specific mean predicted value instead.[22] Standard errors for stratification-based estimators often rely on simplifying assumptions; again, bootstrap may be used as an alternative. The ATT (and ATNT) can similarly be estimated by replacing the ratio $n_j/n$ with a ratio of the stratum proportion of the treated (untreated) population.

Matching is similar in spirit to stratification, but taken to the finest strata: the individual level. For each individual $i$ in the sample, we select $M \geq 1$ individuals, $i'$, who are matched to $i$ based on some matching criterion and matching method. Then the estimators of the ATE and ATT are, respectively,

$$n^{-1} \sum_{i=1}^{n} (2A_i - 1) \left( Y_i - M^{-1} \sum_{i'} Y_{i'} \right) \quad \text{and} \quad (\#\mathcal{I}_{A=1})^{-1} \sum_{i \in \mathcal{I}_{A=1}} \left( Y_i - M^{-1} \sum_{i'} Y_{i'} \right),$$

where $i'$ runs over the set $\mathcal{M}_i$ of individuals matched to $i$.

In practice, the following algorithm should be followed:

1. Choose a matching criterion, $C_{i,i'}$ such as nearest neighbour, the Mahalanobis distance, or vector norm, and implement a matching method given the criterion. The criterion may be applied to $L$ or to a summary such as the propensity score $\hat{e}(L)$.

2. Evaluate the quality of the matched sample by carrying out balancing checks described below.

3. If balance is not satisfactory, return to step 1.

There are several factors to consider in a matched analysis, such as the number of matches per individual, $M$; whether to match with or without replacement; if matching without replacement, whether to use greedy matching or the more computationally intensive optimal matching. A discussion of the relative merits and the impact of these choices on bias and variance can be found in a review by Stuart.[23] If balance remains unsatisfactory or to increase robustness, outcome regression as described in Section 4.1.1 can be performed within the matched sample.

Several standard softwares include packages that implement matching and, in some cases, covariate balance checks. Note that the bootstrap should not be used to compute standard errors following matching, and that suitable standard errors depend on how the matching was carried out (e.g. whether with replacement or not).[24]

### 4.1.3 | Inverse probability weighting

The idea behind inverse weighting is to construct a pseudo-sample in which there are no imbalances on measured covariates between the treatment groups. While IPW can be used for treatments measured only at baseline, its strength is with time-varying treatments. Let $W_i$ be the inverse of the probability of the *received* treatment, defined as $W_i = P(A_i = a_i | L_i)^{-1} = e(L)^{-1}$. Assuming no interference, consistency, NUC, and correct specification of the PS model, the average potential outcome if the whole population were treated can, under causal consistency, be shown to equal

$$E\left[W_i A_i Y_i\right] \tag{1}$$

An alternative definition of the weights, denoted stabilised weight, is $W_i = P(A_i = a_i) P(A_i = a_i | L_i)^{-1}$ and is often preferred as it follows naturally from the theoretical derivation of IPW estimators[25] and for time varying exposures typically leads to less extreme values and more stable estimates.[3] In practice, an estimated propensity score is used in place of $P(A_i = 1|L)$ and $P(A_i = 1)$ is replaced by a simple sample average before an empirical average is taken:

$$\hat{E}[Y_{\mathbf{a}(1)}] = n^{-1} \sum_{i=1}^{n} w_i a_i y_i, \tag{2}$$

where $w_i$ are such estimates of $W_i$. If there are *many* people with a given set of characteristics $l_i$ who are treated, but few with this characteristic who are not treated, then $P(A_i = 1 | L_i = l_i)$ will be 'large' and its inverse 'small' so these treated individuals will be downweighted in the sample.

Similarly, an estimate of the average potential outcome if the whole population were set to be *un*treated is

$$\hat{E}[Y_{\mathbf{a}(0)}] = n^{-1} \sum_{i=1}^{n} w_i (1 - a_i) y_i. \tag{3}$$

As before, if there are *many* people with a given set of characteristics who are treated, but few who are not treated, then $P(A_i = 0 | L = l_i)$ will be 'small' and its inverse 'large' so that these people are upweighted. This approach is well-known in the survey sampling literature,[26] where it is used to adjust for unequal sampling fractions – typically the oversampling of certain smaller but important subgroups in a population. When the weights are extreme, they may be truncated or normalised.[27]

As before, the PS is usually estimated via a parametric model. So, similarly to previously described estimation steps, the IPW estimation procedure is straightforward and consists of:

1. Fitting the PS model: e.g. logistic regression model for the probability of being treated.

2. Calculating the weights:

   (a) Use the fitted PS to predict the probability that a person received the treatment s/he did in fact receive.

   (b) Set each individual's weight to one over the probability computed in (2a). "Stabilize" this weight by including the simple probability of being treated in the numerator. Optionally: truncate weights or normalize weights.

   (c) Check the confounders' balance in the weighted sample. If balance is inadequate, return to step 1.

3. Fitting the outcome model: weighting each individual by the weights computed in (2b), fit a regression model for the outcome given the treatment. The treatment coefficient is an estimate of the ATE.

Following the estimation procedure above, standard errors must be computed analytically or via bootstrap to account for estimation of the weights. Robust or empirical standard errors provide reasonable coverage, although they do not explicitly account for the fitting of the PS model.

To estimate the ATT, rather than the ATE, we change our focus to $E[Y_{\mathfrak{a}(1)} - Y_{\mathfrak{a}(0)}|A = 1]$. Clearly, we can compute an estimate of $E[Y_{\mathfrak{a}(1)}|A = 1]$ with little trouble, as this is easily identified and estimated in the data by

$$\hat{E}[Y_{\mathfrak{a}(1)}|A = 1] = (\#I_{A=1})^{-1} \sum_{i \in I_{A=1}} a_i y_i.$$

The second term, $E[Y_{\mathfrak{a}(0)}|A = 1]$, requires a bit more work: this is an average of the potential outcome $Y_{i,\mathfrak{a}(0)}$ in the (impossible) situation where the $i$ indexes those who were in fact treated. It turns out that we can again use reweighting of the observed sample of the untreated individuals by

$$\hat{E}[Y_{\mathfrak{a}(0)}|A = 1]] = n^{-1} \sum_{i=1}^{n} w_i^{ATT}(1 - a_i)y_i, \tag{4}$$

with stabilized weights equal to

$$w_i^{ATT} = \frac{\hat{P}(A_i = 1|L_i)}{\hat{P}(A_i = 0|L_i)} \times \frac{\hat{P}(A_i = 0)}{\hat{P}(A_i = 1)}.$$

As before, the weighting has been used to construct a pseudo-population in which there are no imbalances on measured covariates between the exposure groups. In the case of the ATT, we do so by re-balancing the distribution of the covariates in the unexposed group only.

Care must be taken as, in practice, a small number of large weights can be highly influential, though this may be mitigated through ad-hoc effective solutions such as shrinking of the largest weights to a smaller value such as the 99th percentile of the weight distribution (often referred to as truncation).

### 4.1.4 | A hybrid approach: doubly robust estimation

Inverse probability of treatment weighted estimators can also be *augmented*. Note that

$$E[Y_{\mathfrak{a}(a)}] = E[Y_{\mathfrak{a}(a)} - \mu_{\mathfrak{a}(a)}(L)] + E[\mu_{\mathfrak{a}(a)}(L)]$$

where here, $\mu_{\mathfrak{a}(a)}(L)$ is the expected outcome with $A$ set to $a$ and covariates taking values $L$. Invoking the consistency and NUC assumptions, we have $\mu_{\mathfrak{a}(a)}(L) = E[Y|A = a, L = l]$ which is, in practice, replaced by a parametric model. This gives rise to the alternative estimator

$$\hat{E}[Y_{\mathfrak{a}(a)}] = \frac{1}{n} \sum_{i=1}^{n} \frac{I[A_i = a](y_i - \hat{\mu}(\mathfrak{a}(a), l_i))}{\hat{P}(A_i = a|L = l_i)} + \frac{1}{n} \sum_{i=1}^{n} \hat{\mu}(\mathfrak{a}(a), l_i), \tag{5}$$

with $I[\cdot]$ the indicator function that takes value 1 when condition $\cdot$ holds and 0 otherwise, and $\hat{\mu}_a(l)$ a model-predicted mean for $Y$ with $A$ set to $a$ and $L$ as observed. The estimator (5) is *doubly robust*, which means that it is consistent even if one of $P(A_i = a|L = l_i)$ and the modelled mean $\mu_{\mathfrak{a}(a)}(L)$ is misspecified. If both models are correctly specified, then the augmented inverse weighted estimator is at least as efficient as the unaugmented inverse weighted estimator.

Bang & Robins,[28] building on Scharfstein et al.,[29] reformulated the augmented estimator, noting that it can be viewed as an unweighted regression that includes the inverse of the propensity score as a covariate. Thus, what first appears to be a somewhat complicated estimator is in fact very easy to implement. Unlike for the propensity score model, the outcome regression model in augmented inverse probability weighted estimation can rely on conventional variable selection approaches.

## 4.2 | Instrumental variable based methods

All methods described so far yield valid estimates under the NUC assumption. This assumption is easily violated in observational studies, where the prognosis of patients tends to determine the choice of treatment and the reasons for a specific treatment choice are seldom completely registered or, more generally, the exposure level is influenced by unmeasured factors. One alternative approach is instrumental variable (IV) analysis which can handle both measured and unmeasured confounding. Unbiased estimation results once a 'pseudo random variable' or so called 'instrumental variable' is identified *and* some additional assumptions hold. The method originates from econometrics[30,31] and is becoming increasingly popular in medical research. The literature on IV, with examples, is vast.[32,33,34,35,36] We will discuss here the general IV assumptions, typical causal estimands, and the corresponding estimation procedures that are most commonly used. To focus on the principles here, our formalization below ignores the measured baseline covariates $L$, although the approach extends quite naturally to conditioning on them.

An IV analysis aims to resemble that of a RCT, by using one or more variables (instruments) associated with treatment, but not in any other way related to the outcome. The instrument can be seen as a surrogate for randomisation. This is depicted in Figure 2 where $Z$, the instrumental variable, is associated with $A$ (the Figure suggests a causal relation but that is not necessary, association is sufficient). The instrument $Z$ is related to response $Y$ only via the treatment $A$; and the instrument is independent of unmeasured confounders $U$.

Instrumental variable analysis can be used in trials to study the effect of non compliance,[30,37,38] as in our BEP example, where randomisation to the offer of the breastfeeding programme could be used as instrument for attending the programme. Variation in preference for a certain treatment among physicians[39,40] or variation in treatment policies among medical centers[41] are other examples of variables which can be considered close to pseudo-randomization for treatment or policy assignment. When physicians have strong preferences for one or another treatment, identical patients may receive different treatments; a variable measuring the physician's preference, like the percentage of prescriptions $A = 1$ in a certain time window, could be used here as an instrument. Another popular IV approach is found in so called 'Mendelian randomisation' studies where genetic variation takes the role of the instrumental variable.[42,43]

### 4.2.1 | The three core IV assumptions

To be an instrumental variable for the causal effect of $A$ on $Y$, $Z$ should satisfy the following three core assumptions (possibly conditional on $L$):

IV1 $Z$ is associated with the treatment $A$ of interest;

IV2 $Z$ is independent of any unmeasured confounders of the $A \rightarrow Y$ relationship;

IV3 $Z$ is independent of the outcome $Y$ conditional on treatment $A$ and unmeasured confounders $U$.

Unfortunately only assumption IV1 can be empirically checked in the data.[44] Assumptions IV2 and IV3 are not verifiable in the data: only their plausibility can be examined. For example the observation that $Z$ is independent of all observed confounders, makes assumption IV2 more plausible. Situations in which these assumptions are likely or unlikely to hold are discussed for Mendelian randomisation and for physician's preference by several authors.[42,43,32] Assuming $Z$ is an IV and that no interference, consistency and positivity hold, IV-based estimation does not require the NUC assumption to lead to an estimator. However, an IV estimator on its own can only provide bounds for causal treatment effect.[45,46] These bounds are generally so wide that they are not useful. In order to obtain point estimates, additional assumptions are needed as discussed below.

### 4.2.2 | Standard IV estimation

There are several ways of obtaining point estimates in an IV analysis. The traditional IV estimator is the Wald estimator,[47] which dates back to Write's publication in 1928,[48] and equals:

$$\hat{\beta}_{IV} = \frac{\widehat{cov}(Y, Z)}{\widehat{cov}(A, Z)} = \frac{\sum_{i=1}^n y_i z_i - \frac{1}{n}(\sum_{i=1}^n y_i)(\sum_{i=1}^n z_i)}{\sum_{i=1}^n z_i a_i - \frac{1}{n}(\sum_{i=1}^n z_i)(\sum_{i=1}^n a_i)}.$$

This estimator is based on two relationships which can be estimated unbiasedly because they are unconfounded: the relationship between instrument $Z$ and outcome $Y$, and the relationship between instrument $Z$ and treatment $A$. In case of a binary instrument, this expression reduces to

$$\hat{\beta}_{IV} = \frac{\hat{E}[Y | Z = 1] - \hat{E}[Y | Z = 0]}{\hat{E}[A = 1 | Z = 1] - \hat{E}[A = 1 | Z = 0]}, \tag{6}$$

where $\hat{E}[Y | Z = z]$ refers to the simple average of $Y$ in the selected subset with $Z = z \in \{0, 1\}$. Similarly, $\hat{E}[A | Z = z]$ is a simple average of $A$ in the selected subset with $Z = z$.

The numerator expresses the effect of instrument on outcome; the mean difference between those with $Z = 1$ and $Z = 0$, or the risk difference in the case of a binary outcome. To obtain an estimate of the effect of the treatment on the outcome, the effect of the instrument on the outcome is inflated by dividing the numerator by the effect of the instrument on the treatment. The smaller the correlation between $Z$ and $A$ (the so-called strength of the instrument), the larger the inflation factor.

The traditional IV estimator can be equivalently obtained through a two stage linear regression (2SLS) approach. In the first stage, a linear regression model is fitted with treatment $A$ as dependent variable and the instrument $Z$ as an independent variable (and optionally measured confounders $L$), yielding for each subject $\hat{E}[A | Z = z_i]$. In the second stage, a linear regression model is fitted to the outcome $Y$ on $\hat{E}[A|Z]$ (and possibly $L$). The regression coefficient for $\hat{E}[A|Z]$ is the IV estimate of the treatment effect, under assumptions which we will discuss next.

Many authors apply 2SLS methods to binary outcomes by fitting linear regression outcome models and hence yielding estimates of risk differences. This is not advisable when also including covariates $L$ as the fitted model may predict outcome values $> 1$ or $< 0$. Extending the two stage approach to a logistic regression outcome model would be hampered by the non linearity of the logistic model. A two stage approach with a linear model in the first stage and a logistic model in the second stage can only be used to obtain IV estimates of odds ratios if the outcome is rare. Otherwise, it is preferable to use logistic structural mean models.[49,50,51]

### 4.2.3 | Assumptions needed for a causal effect interpretation

The three main IV assumptions are not sufficient for a causal interpretation of the IV estimator. There are several options for additional assumptions, which may lead to different causal interpretations of the IV estimator.

1. The most strict assumption demands that the effect of the treatment is the same within levels of $U$, by assuming that $E[Y|A, U] = \beta_0 + \beta_1 A + \beta_2 U$.[37,49] Then the 2SLS IV estimator consistently estimates the ATE. Assuming the same treatment effect for all individuals is, in general, unrealistic as more severely affected patients may benefit more (or less) from treatment, the treatment could interact with other drugs, men and women could respond differently, and so on.

2. Assumptions regarding a homogeneous treatment effect can be relaxed using *structural mean models* (SMM).[52,49] An SMM is a model for the mean difference between an observed outcome $Y$ and a potential outcome such as $Y_{\mathfrak{a}(0)}$, that may condition on observed treatment $A$ and instrument $Z$. A simple SMM is:

$$E[Y - Y_{\mathfrak{a}(0)}|A, Z] = A\beta_1, \tag{7}$$

which assumes that $E[Y - Y_{\mathfrak{a}(0)}|A, Z]$ does not depend on $Z$. Furthermore for $A = 0$ we have $E[Y - Y_{\mathfrak{a}(0)}|A = 0, Z] = 0$, which is exactly the (mean) consistency assumption for $\mathfrak{a}(0)$. When $A = 1$ we obtain $E[Y - Y_{\mathfrak{a}(0)}|A = 1, Z] = \beta_1$, and if the consistency assumption holds $\beta_1$ equals the ATT. Simple unbiased estimating equations can be defined from this model,[37] with solution equal to the Wald estimator. Baseline covariates $L$ could be included in this model, including interactions between $L$ and exposures $A$.[37]

   If one additionally assumes 'no current treatment effect interaction', i.e. the structural mean effect of treatment does not depend on the observed treatment, then this expression can also be used to calculate the $E[Y_{\mathfrak{a}(1)}|L]$ and hence the ATE.

3. An alternative assumption is *monotonicity*, which we will explain here for the situation with a binary instrument $Z$ that causally affects a binary treatment $A$. With $A_{\hat{\mathfrak{z}}(z)}$ the value of $A$ when $Z$ is set to $z \in \{0, 1\}$, four types of subjects can be identified:

   (1) always takers: those with $A_{\hat{\mathfrak{z}}(1)} = A_{\hat{\mathfrak{z}}(0)} = 1$, i.e. subjects who will use treatment regardless of the value of the instrument;

   (2) never takers: those with $A_{\hat{\mathfrak{z}}(1)} = A_{\hat{\mathfrak{z}}(0)} = 0$;

   (3) compliers: those with $A_{\hat{\mathfrak{z}}(1)} = 1$ and $A_{\hat{\mathfrak{z}}(0)} = 0$; and

   (4) defiers: those with $A_{\hat{\mathfrak{z}}(1)} = 0$ and $A_{\hat{\mathfrak{z}}(0)} = 1$.

   The monotonicity assumption implies that defiers do not exist,[30] i.e. $A_{\hat{\mathfrak{z}}(1)} \geq A_{\hat{\mathfrak{z}}(0)}$. Under this assumption, the IV estimator (6) consistently estimates a local causal effect, i.e. a causal (average) treatment effect in the subgroup of compliers, or the Complier Average Causal Effect (CACE):

$$CACE = E[Y_{\mathfrak{a}(1)} - Y_{\mathfrak{a}(0)}|A_{\hat{\mathfrak{z}}(1)} = 1 \text{ and } A_{\hat{\mathfrak{z}}(0)} = 0)]$$

   The interpretation of the CACE is often difficult,[48,53,54] because the subgroup of compliers cannot be identified from the data, although general characteristics like the distribution of age and sex can be obtained.[55] In some particular instances, however, it could be the parameter of interest: in our BEP example, the CACE represents the effect of the programme in the subgroup of individuals for which the programme is acceptable and accepted, e.g. the CACE for $A_2$ is the effect among those individuals who will attend the breastfeeding programme when invited but not otherwise. Although this formulation is appealing, when the instrumental variable is continuous, defining monotonicity is more complicated and the interpretation often even less intuitive.[56,54]

### 4.2.4 | When are IV methods useful?

We have discussed the IV assumptions needed to estimate causal treatment effects. Although many IV estimators are consistent, in finite samples instrumental variable estimators are generally biased. The bias depends on the sample size and on the strength of the instrument (i.e the correlation between $Z$ and $A$).[57] Furthermore, IV estimates are very sensitive to deviations from the IV assumptions. A small association between the unmeasured confounders and the instrument can lead to substantial bias especially if the instrument is weak.[57,58] Moreover, weak instruments yield very imprecise IV estimates and often (very) large sample sizes are needed to obtain informative results.[59] This implies that instruments should be strongly correlated to the treatment. There is however a trade-off between the amount of unmeasured confounding and the strength of the instrument: an instrument cannot be strong if there is substantial unmeasured confounding[58] and a strong instrument implies weak unmeasured confounding.

To summarize, an instrumental variable analysis may be useful in the following situations: (1) the amount of expected unmeasured confounding is substantial, (2) an instrument exists for which the core IV assumptions are plausible and additionally a fourth assumption to interpret the point estimate can be sensibly invoked, (3) the instrument is sufficiently strong, and (4) sample sizes are sufficiently large (when instruments are weak, required sizes may be in the order of several thousands of subjects). Otherwise methods assuming NUC should be considered, while also maximizing the number of measured confounders. Although approaches relying on NUC yield biased estimators if unmeasured confounding is present, the direction of the bias is often known and the size of the bias may be approximated in sensitivity analyses.

## 4.3 | Choosing an estimation method

Table 3 reviews several points that go to the heart of which causal estimands are meaningful and relevant in the specific setting represented by our case study. An accompanying Table 4 summarises the main assumptions that are invoked by the various methods reviewed in this section when aiming to estimate the ATE (in addition to no interference and causal consistency). The Table is self-explanatory and highlights that the core difference lies in whether we are prepared or not to assume NUC, given a vector of measured confounders $\boldsymbol{L}$. However it is worth stressing these additional points.

For those methods assuming NUC:

- Outcome regression assumes a correct specification of the outcome model

- PS-matching and PS stratification assume that the PS balances the confounder distribution

- IPW assumes that the PS model is correctly specified given a sufficient set of confounders.

- Linear outcome regression models that condition on the estimated PS, as opposed to the original vector of confounders $\boldsymbol{L}$, require that either the outcome model or the PS model is correct and that the treatment effect does not vary with the PS.[60]

- The specification of the PS model should achieve balancing of the distribution of the measured confounders across treatment arms. Achieving this aim is substantially different from achieving treatment prediction, and hence the criteria used for the latter do not apply here.

- In general, outcome regression is more efficient than a PS-based method.

- The choice between PS-based methods (i.e. stratifying, regression adjustment, matching and IPW) depends on whether efficiency is an issue. Weighting may be inefficient (unless a doubly robust approach is used) if there are subjects with a very high or low PS value; matching has a trade off between a close match (which implies loss of efficiency because not all subjects are matched) versus residual confounding.

When not assuming NUC

- IV estimation replaces the NUC assumption with other rather stringent assumptions.

- IV methods yield estimates that are very inefficient when instruments are weak and suffer from small sample bias.[57]

With any given approach come choices in implementation that imply a trade-off between bias and variance. For example, in the context of PS matching, the use of smaller calipers to determine a match will reduce bias but may lead to a smaller matched sample and hence loss in efficiency. In PS-inverse weighting, the use of weight truncation to reduce the influence of a small number of points has the effect of decreasing the variance at the cost of introducing some bias. It is hence impossible to recommend a single "best" approach, but rather choices are specific to the context where researchers must balance bias, statistical efficiency, and in some cases computational efficiency.

## 5 | RESULTS FROM THE SIMULATION LEARNER

We applied the methods discussed in the previous section to estimate the ATE and the ATT of $A_1$, $A_2$ and $A_3$ on weight at 3 months using the data from the simulation learner PROBITsim. More details and the code used to produce the reported results are given in Appendix 2 and in the material available at www.ofcaus.org.

### 5.1 | Effect of the randomized programme offer ($A_1$)

First we estimate the causal effect of the randomized offer of the BFE ($A_1$) on weight at 3 months. This is simply the difference in mean weight at 3 months between those with $A_1 = 1$ and $A_1 = 0$ because $A_1$ is randomized. This is also an estimate of the intention-to-treat (ITT) effect, in this case an "intention to educate", and is most relevant for health policy makers. This estimate is 94.2 grams (95% confidence interval: 76.4 to 112.0 grams). It indicates that inviting all expecting mothers in the study population to attend this specific programme would increase their baby's weight by 94 grams, on average. The true value obtained from Table 2 was 98 grams and is well within the confidence interval.

### 5.2 | Effect of programme uptake ($A_2$)

Table 5 shows the estimated ATE for $A_2$, which is the effect most directly relevant to women deciding whether or not to attend the programme if offered. We also show the corresponding estimated ATT. In Section 3 we showed that the true $ATE_2$ was greater than $ATT_2$ (165.1 vs. 152.8 grams), whereby the treated, i.e. the mothers who attended the programme, were on average, more educated and their infants had higher weight at 3 months but smaller increases from attending the programme. We estimated these target parameters under different assumptions and model specifications, starting from crude estimates where confounding is ignored ($\widehat{ATE}_2$=196.0 grams and $\widehat{ATT}_2$=148.7 grams). We then controlled for measured confounding via outcome regression, adopting two alternative model specifications that included all the potential confounders for the $A_2$ to weight at 3 months relationship: maternal age, education, allergy status, smoking during pregnancy and area of residence. In the first specification we included a quadratic term for maternal age, and in the second we also included interactions between $A_2$ and each confounder. The first led to $\widehat{ATE}_2$=155.4 grams and the second to $\widehat{ATE}_2$=165.0 grams, much closer to the true value of 165.1 grams.

When applying the PS-based methods, we fitted the PS model by logistic regression with the same confounders (including the quadratic term for maternal age). Stratification (over 6 strata) led to the same estimates as the more general outcome regression models ($\widehat{ATE}_2$=165.0 and $\widehat{ATT}_2$=148.7 grams), while matching, either to 1 or 3 other infants, led to slightly smaller and less precise, estimates. Balance checks revealed that the PS model was well specified (see Appendix 2). Adopting inverse weighting or doubly robust estimation gave estimates closer to those from outcome regression, also in terms of precision.

The reported IV estimate used $A_1$ as the instrument and assumed no $A_1$–$A_2$ interaction to be interpreted as an ATE. This was estimated at 146.2 grams and, as expected, has a very large estimated standard error.

### 5.3 | Effect of starting breastfeeding ($A_3$)

The estimated ATE and ATT for the effect of $A_3$ on infant weight at 3 months are found in Table 6 . As before they are obtained under different assumptions and using different methods. As their true values depend on whether the exposure is set in a world where the BEP is or not present, results are reported separately under these two scenarios.

Note also that the true average potential outcome in the world where no programme was offered but all mothers start breastfeeding was lower than in the world where BEP is offered to all mothers and they all start breastfeeding (**??**, rows 8 and 9) because of the effect of the BEP on breastfeeding duration. This impacts on the causal effect of breastfeeding when $A_1$ is set at 0, $ATE_{3,\mathbf{a}_1(0)} = 386.8$ grams and $ATT_{3,\mathbf{a}_1(0)} = 380.1$ grams; when $A_1$ set to 1, $ATE_{3,\mathbf{a}_1(1)} = 422.3$ grams and $ATT_{3,\mathbf{a}_1(1)} = 421.4$ grams.

The confounders of the $A_3$ to weight at 3 months relationship include not only maternal age, education, allergy status, smoking during pregnancy and area of residence (i.e. those involved in the analyses of $A_2$), but also the infant's sex, birth weight (including a quadratic term) and whether the infant was born by caesarean section. In the analyses concerning the world where $A_1$ is set to be 1, $A_2$ is also a confounder as it influences both $A_3$ and infant weight.

There is little difference across the ATE estimates, obtained using either outcome regression or PS-based methods: the results are all very similar and standard errors, while variable, all still lead to the conclusion that $A_3$ meaningfully and statistically affects the outcome. Balance checks for these two scenarios revealed that the PS model was relatively well specified in both, although there was greater separation (less overlap) than for PS-model for $A_2$ (see Appendix 2).

We do not produce an equivalent IV estimate as there is no suitable IV for this effect, since $A_1$ violates the second IV assumption: $A_1$ influences the outcome not only via $A_3$ but also via $A_2$.

For the ATT estimates, regression adjustment seems to perform better than the other methods, especially in the world where $A_1 = 1$. Of course, our simulation learner has generated just one relatively simple world model where both our outcome and propensity model are easy to specify.

## 6 | DISCUSSION

We set out to discuss 'the making of' a causal effect question involving a well-defined point exposure for which we seek to find the average treatment effect, possibly conditional on baseline characteristics. We have maintained an emphasis on the framing of the scientific causal question, and in considering many methods together, in their basic form, so as to compare and contrast the required assumptions of different principled estimation approaches for directly targeted estimands.

We applied the concept of principled estimation in turn to four different options for exposure levels which present themselves along the path from treatment prescription to completion. As we moved with the selected exposure along this path, the sufficient set of baseline confounders (and effect modifiers) became richer, and we had to account for what happened earlier in the path. In doing so, we saw that we cannot treat randomization as 'once an instrument, always an instrument'. Rather, randomization (our $A_1$) may act as an instrument for the effect of following the programme ($A_2$), but it violated the assumptions required for it to be an instrument when studying the effect of 'starting breastfeeding' ($A_3$). At every instance, thought is required to adapt to the new situation and estimate a relevant causal effect in a (sub)population of interest.

In a similar vein, confounders that act as effect modifiers could be conditioned on to estimate average causal effects within specific population strata. Subsequently, we can average over their distribution in the population of interest. With additional averaging, we lose some ability to offer stratified evidence and provide personalized information but uphold a more global public health perspective. This pertains to both the ATE and ATT target.

For selected estimands, we showed how the various estimating approaches perform in their most basic form. We recognized that many of them operate under similar identifying assumptions. For example the different propensity methods all assume correct specification of the PS model, and when choosing one of the methods one should consider additional issues. For the stratification, the choice of the number of strata and residual bias, for the matching the trade of between finding matched individuals and the fineness of the matching, and for the inverse probability weighting, the size of the weights, truncation. Of course, differences remain in operating characteristics when key (untestable) assumptions are violated. The list of available approaches under the NUC assumption includes familiar standardized means derived from the classic regression of outcome on baseline covariates and the exposure. This need not perform worse, and can even be better than more novel PS-based methods that seek covariate balance after using the propensity score for regression, matching, stratification or inverse probability weighting. Doubly robust methods may be expected to outperform others when one set of model assumptions is violated, but equally loses precision (increases error) when both the outcome regression and PS model are ill-fitting, and may be inefficient in finite samples when only one model is correctly specified.[61]

When we cannot find a sufficient set of confounders, instrumental variable approaches form an appealing alternative provided an instrument can be found. To interpret the resulting estimator additional assumptions are needed that are not always easy to justify; whether those can lead to very broad confidence intervals. There are other alternative routes still, such as regression discontinuity designs for instance,[62] a variation on pseudo-randomization that is found in specific designs.

We set out to give an overview of the basic principles that guide causal inference however in practice, many complications conspire to challenge the applied statistician when performing causal inference. We, for instance, have implicitly assumed all covariates are measured without error and there is no selection bias or drop-out. In practice, data may be not just confounded, but may also suffer from missingness[28] and measurement error on exposure[63] or confounders[64] is likely. Flexible models may be more appropriate to capture the associations involved. Clustering and non-interference may require extension of the presented set-up to incorporate interference.[65,12,66] With substantial dropout from a longitudinal outcome due to mortality, one must adapt the definition of the outcome explicitly or reduce the target population to potential survivors on all treatments considered.[67] In the international initiative of Strengthening Analytical Thinking for Observational Studies (STRATOS)[68] other topic groups focus on guidance for these topics and joint developments with our causal inference topic group are envisaged for the future.

We have purposefully focused on the point (i.e. fixed) exposure perspective, even though we considered a natural sequence of exposures. This allowed us to present an overview of different estimation principles, showing how they resemble one another, and where and how they differ in their fundamental assumptions and performance. The natural next step is to consider the joint effect of a sequence of exposure options $a_2, a_3, a_4$ as a time-varying treatment regime and engage in estimating causal effects of different (static or dynamic) treatment strategies. To achieve this, we would need to formally account for time-varying confounders along that path (see, for example[69,70]). We might further aim to explain the total effect and engage in mediation analysis to evaluate the possible role of intermediate variables on the causal path.[71,60,72] For all these endeavours in higher dimensions, the principles laid out here continue to form an important foundation.

Even at the point exposure level, the literature on adaptations of these estimators under additional or alternative assumptions is vast, but beyond the scope of this tutorial. Here, we focused on a binary exposure and a continuous, uncensored outcome. When exposures are categorical or continuous, a *generalized* propensity score can be used.[73,74]

In the course material that accompanies this paper, practical exercises discuss estimation when the primary outcome is binary, using the Right Heart Catheterisation dataset.[75] Estimating a linear effect, a risk difference, is less obvious there and requires extra care.

We hope the lay-out of this principled approach will inspire practising statisticians to think carefully about what they are estimating and to report as clearly as possible on the nature of their exposure and causal estimand, as well as the assumptions on which they have relied. A naive analysis can be dangerous when followed by either implicit or explicit causal claims that are made without regard for confounding or effect modification or for their population level interpretation. We hope this contribution can generate confidence and insight into methodological ground-rules, and promote better thinking, reliable estimates and clear reporting.

## Acknowledgements

This work was developed on behalf of Topic Group Causal inference (TG7) of the international initiative of Strengthening Analytical Thinking for Observational Studies (STRATOS). The objective of STRATOS is to provide accessible and accurate guidance in the design and analysis of observational studies. The authors thank the Lorentz Center Leiden, for the opportunity to organise a STRATOS workshop and the members of the workshop, the STRATOS publication panel and Vanessa Didelez for their comments on an earlier version of this paper. E.M.M. Moodie acknowledges the support of the Natural Sciences and Engineering Research Council (NSERC) of Canada, Discovery Grant #RGPIN-2014-05776 and a Chercheur-boursier senior career award from the Fonds de recherche du Québec, Santé. I. Waernbaum acknowledges the Swedish Research Council grant # 2016-00703. B. De Stavola acknowledges UK Medical Research Council Grant # MR/R025215/1

**TABLE 1** A selection of causal estimands for exposures $A_1$ and $A_2$

| Estimand | Definition |
|---|---|
| **Effect of Programme Offer ($\mathfrak{a}_1$)** | |
| ATE$_1$=ATT* | Average Treatment Effect |
| | $E[Y_{\mathfrak{a}_1(1)}] - E[Y_{\mathfrak{a}_1(0)}]$ |
| **Effect of Programme Uptake ($\mathfrak{a}_2$)** | |
| ATE$_2$ | Average Treatment Effect |
| | $E[Y_{\mathfrak{a}_2(1)}] - E[Y_{\mathfrak{a}_2(0)}]$ |
| ATT$_2$ | Average Treatment effect among the Treated† |
| | $E[(Y_{\mathfrak{a}_2(1)}|A_2 = 1, A_1 = 1] - E[Y_{\mathfrak{a}_2(0)}|A_2 = 1, A_1 = 1]$ |
| ATNT$_2$ | Average Treatment effect among the Non-Treated† |
| | $E[Y_{\mathfrak{a}_2(1)}|A_2 = 0, A_1 = 1] - E[Y_{\mathfrak{a}_2(0)}|A_2 = 0, A_1 = 1]$ |

\* Intention to treat

† Note that the ATT and ATNT for $\mathfrak{a}_2$ can only be derived from the (random) subgroup $A_1 = 1$ since the the programme is only available within the randomized trial and to those assigned to it being offered.

**TABLE 2** True average potential infant weight at 3 months under different interventions in different (sub)populations.

| Potential outcome | Interventions | overall | $A_2=1$ | $A_1=1$ $A_2=0$ | $A_1=1$ $A_3=1$ | $A_1=1$ $A_3=0$ | $A_1=0$ $A_3=1$ | $A_1=0$ $A_3=0$ | Education low | int | high |
|---|---|---|---|---|---|---|---|---|---|---|---|
| $Y_{\mathbf{a}_1(0)}$ | BEP not offered | 6017 | 6047 | 5964 | 6149 | 5733 | 6274 | 5761 | 5914 | 6057 | 6141 |
| $Y_{\mathbf{a}_1(1)}$ | BEP offered | 6115 | 6200 | 5964 | 6292 | 5733 | 6308 | 5923 | 6024 | 6155 | 6207 |
| $Y_{\mathbf{a}_2(0)}$ | BEP not followed | 6017 | 6047 | 5964 | 6149 | 5733 | 6274 | 5761 | 5914 | 6057 | 6141 |
| $Y_{\mathbf{a}_2(1)}$ | BEP followed | 6182 | 6200 | 6149 | 6308 | 5911 | 6329 | 6035 | 6128 | 6208 | 6226 |
| $Y_{\mathbf{a}_1(1),\mathbf{a}_3(0)}$ | | | | | | | | | | | |
| $Y_{\mathbf{a}_2(0),\mathbf{a}_3(0)}$ | no BF | 5827 | 5849 | 5788 | 5871 | 5733 | 5893 | 5761 | 5730 | 5854 | 5981 |
| $Y_{\mathbf{a}_1(0),\mathbf{a}_3(1)}$ | BEP not offered, BF started | 6214 | 6226 | 6193 | 6251 | 6133 | 6274 | 6153 | 6154 | 6248 | 6246 |
| $Y_{\mathbf{a}_1(1),\mathbf{a}_3(1)}$ | BEP offered, BF started | 6249 | 6282 | 6193 | 6292 | 6157 | 6308 | 6191 | 6207 | 6276 | 6262 |
| $Y_{\mathbf{a}_2(1),\mathbf{a}_3(1)}$ | BEP followed, BF started | 6277 | 6282 | 6270 | 6308 | 6212 | 6329 | 6225 | 6261 | 6292 | 6266 |
| $Y_{\mathbf{a}_4(1)}$ | duration BF = 3 month | 6351 | 6345 | 6362 | 6372 | 6307 | 6392 | 6311 | 6393 | 6339 | 6286 |

BEP: Breastfeeding encouragement programme; BF: breastfeeding; int: intermediate

$A_2=1$: women who followed the breastfeeding programme

$A_2=0$ and $A_1=1$: women who were offered the breastfeeding programme but did not follow it

$A_3=1$ and $A_1=1$: women who started breastfeeding in the intervention group

$A_3=1$ and $A_1=0$: women who started breastfeeding in the control group

$A_3=0$ and $A_1=1$: women who did not start breastfeeding in the intervention group

$A_3=0$ and $A_1=0$: women who did not start breastfeeding in the control group

$Y_{\mathbf{a}_1(0)}$: the potential outcome that would occur if $A_1$ were set to take the value 1. Similar for $Y_{\mathbf{a}_1(1)}, Y_{\mathbf{a}_2(1)}, Y_{\mathbf{a}_3(0)}Y_{\mathbf{a}_4(1)}$

$Y_{\mathbf{a}_1(0),\mathbf{a}_3(1)}$: The potential outcome under a double intervention with $A_1$ set to 0 and $A_3$ set to 1. Similar for $Y_{\mathbf{a}_1(1),\mathbf{a}_3(1)}, Y_{\mathbf{a}_2(1),\mathbf{a}_3(1)}$

Results for $Y_{\mathbf{a}_1(0)}$ and $Y_{\mathbf{a}_2(0)}$ are equal, because BEP only affects the outcome if the programme is followed

Results for $Y_{\mathbf{a}_1(0),\mathbf{a}_3(0)}, Y_{\mathbf{a}_1(1),\mathbf{a}_3(0)}, Y_{\mathbf{a}_2(1),\mathbf{a}_3(0)}$ are equal because BEP only affects Y via $A_3$ and duration of breastfeeding, if started

The effect of three full months of breastfeeding is not affected by BEP

**TABLE 3** Considerations for the ATE for exposures $A_1,...,A_4$; the same issues arise in estimation of the ATT and ATNT. ITT = Intention-to-treat.

| Exposure | Estimand | Comments |
|---|---|---|
| $A_1$ | ITT effect | Randomization ensures unbiased estimation using simple contrasts. |
| $A_3$ | ATE$|A_1=1$, or ATE$|A_1=0$ | Effect of starting breastfeeding in a world where all (or no) women are offered the programme. If we do not condition on $A_1$, then we mix the two populations (or two "worlds"), which would never co-exist outside of a trial where only half of women are offered the intervention. Further, $A_2$ is an effect modifier. Thus, correct specification of the outcome model requires an $A_2A_3$ term, and the ATE must then marginalize over the distribution of $A_2$. Note that the conditioning on $A_1$ is not relevant for estimating the causal effect of $A_2$, as $A_1$ has the role of an instrument for $A_2$, but not for $A_3$ or indeed for $A_4$. |
| $A_4$ | ATE$|A_3=1$ | There is no support in the data for an effect of $A_4$ in women with $A_3=0$. Note also that $A_4=0$ is a mixture of durations of breastfeeding, potentially from 1 day up to just shy of 3 months. The consistency assumption implies that its estimated effect refers to settings with the same distribution of breastfeeding discontinuation times. An equivalent statement holds for the interpretation of $A_3=0$ in the row above. |

**TABLE 4** Sufficient assumptions for estimation methods of the ATE of a binary single point exposure $A$ (assuming throughout that no interference and consistency hold).

| Method | Assumptions | | | | |
|---|---|---|---|---|---|
| | NUC | Correct specification of | | Core IV assumption | No $Z-A$ interaction |
| | | $Y$ model | $PS^{(*)}$ model | | |
| Outcome regression | | | | | |
| conditional on $L$ | $\surd$ | $\surd$ | | | |
| conditional on $PS = e(L)$ | $\surd$ | $\surd^{(*)}$ | $\surd^{(*)}$ | | |
| Stratification by $e(L)$ | $\surd$ | | $\surd$ | | |
| Matching by $e(L)$ | $\surd$ | | $\surd$ | | |
| IPW by $e(L)$ | $\surd$ | | $\surd$ | | |
| DR via $L$ and $e(L)$ | $\surd$ | either | or | | |
| IV $Z$ | | | | $\surd$ | $\surd$ |

$^{(*)}$ Either of these if the outcome model is linear.

**TABLE 5** Estimated ATE and ATT of $A_2$ on weight at 3 months (in grams) obtained using alternative estimation methods controlling for relevant confounders*; PROBITsim Study.

| Estimand | Estimation method | Estimate | (SE) |
|---|---|---|---|
| ATE | | | |
| | True value | 165.1 | |
| | Crude regression | 196.0 | ( 9.6) |
| | Regression adjustment (without interactions) | 155.4 | ( 9.5) |
| | Regression adjustment (with interactions) | 165.0 | ( 9.7) |
| | PS stratification$^\dagger$ (6 strata) | 165.0 | ( 9.4) |
| | Regression with PS $^\dagger$ | 156.2 | ( 9.0) |
| | PS matching (1 match)$^\ddagger$ | 155.7 | ( 10.1) |
| | PS matching (3 matches)$^\ddagger$ | 154.9 | ( 10.1) |
| | PS IPW$^\dagger$ | 164.7 | ( 9.3) |
| | PS DR IPW$^\dagger$ | 164.7 | ( 9.7) |
| | IV | 146.2 | ( 14.0) |
| ATT | | | |
| | True value | 152.8 | |
| | Regression adjustment (with interactions) | 148.7 | ( 9.4) |
| | PS stratification$^\dagger$ (6 strata) | 148.7 | ( 9.6) |
| | PS matching (1 match)$^\ddagger$ | 145.8 | ( 9.8) |
| | PS matching (3 matches)$^\ddagger$ | 145.4 | ( 9.7) |
| | PS IPW$^\dagger$ | 148.0 | ( 9.6) |

* The variables controlled for in each of these analyses were: maternal age (linear and quadratic term), maternal education, maternal allergy status, smoking status in the first trimester (i.e. before programme allocation), and area of residence.
$^\dagger$ SE estimated by bootstrap with 1,000 replications.
$^\ddagger$ SE estimated according to Abadie and Imbens (2012), assuming that the conditional outcome variance is homoscedastic, i.e. does not vary with the covariates or treatment. This is implemented in Stata with the option `vce(iid)`. This assumption can be relaxed using the option `vce(robust, nn(2))` for the 1 match analysis and `vce(robust, nn(4))` for the 3 matches analysis.

**TABLE 6** Estimated ATE and ATT of $A_3$ on weight at 3 months (in grams) obtained using alternative estimation methods controlling for relevant confounders* and stratified by whether mothers were offered the BFE programme; PROBITsim Study.

| Estimand | Estimation method | $A_1 = 0$ Estimate | (SE) | $A_1 = 1$ Estimate | (SE) |
|---|---|---|---|---|---|
| ATE | | | | | |
| | True value | 386.8 | | 422.3 | |
| | Crude regression | 503.2 | ( 11.6) | 582.0 | ( 12.2) |
| | Regression adjustment (without interactions) | 384.3 | ( 2.8) | 428.0 | ( 3.3) |
| | Regression adjustment (with interactions) | 384.7 | ( 3.2) | 425.3 | ( 2.7) |
| | Regression with PS [†] | 384.4 | ( 3.2) | 425.9 | ( 3.3) |
| | PS stratification[†] (6 strata) | 392.2 | ( 4.1) | 442.0 | ( 6.5) |
| | PS matching (1 match)[‡] | 386.5 | ( 13.7) | 429.0 | ( 17.4) |
| | PS matching (3 matches)[‡] | 380.7 | ( 12.4) | 437.2 | ( 15.2) |
| | PS IPW[†] | 384.7 | ( 3.8) | 426.6 | ( 7.1) |
| | PS DR IPW[‡] | 384.8 | ( 4.0) | 426.7 | ( 7.3) |
| | | | | | |
| ATT | | | | | |
| | True value | 380.1 | | 421.4 | |
| | Regression adjustment (with interactions) | 378.0 | ( 2.9) | 421.7 | ( 2.5) |
| | PS stratification[†] (6 strata) | 388.8 | ( 4.8) | 438.3 | ( 9.5) |
| | PS matching (1 match)[‡] | 384.3 | ( 15.8) | 435.6 | ( 21.2) |
| | PS matching (3 matches)[‡] | 387.9 | ( 13.5) | 441.2 | ( 18.0) |
| | PS IPW[†] | 381.9 | ( 5.3) | 429.2 | ( 10.1) |

* The variables controlled for in each of these analyses were: maternal age (linear and quadratic term), maternal education, maternal allergy status, smoking status in the first trimester (i.e. before programme allocation), area of residence, baby's birth weight (linear and quadratic term), whether birth was by caesarian section and, in the analyses restricted to $A_1 = 1$, whether the mother attended the programme.

[†] SE estimated by bootstrap with 1,000 replications.

[‡] SE estimated according to Abadie and Imbens (2012), assuming that the conditional outcome variance is homoskedastic, i.e. does not vary with the covariates or treatment. This is implemented in Stata with the option `vce(iid)`. This assumption can be relaxed using the option `vce(robust, nn(2))` for the 1 match analysis and `vce(robust, nn(4))` for the 3 matches analysis.